\documentclass[apj]{emulateapj}

\begin{document}

\title{Galaxy Distribution as a Probe of the Ringlike Dark Matter Structure in 
the Galaxy Cluster Cl~0024+17}

%

\author{Bo Qin$^{1}$, Huan-Yuan Shan$^{1}$, and Andre Tilquin$^{2}$}

\affiliation{$^{1}$National Astronomical Observatories, Chinese Academy of
Sciences, Beijing 100012, China\\
$^{2}$Centre de Physique des Particules de Marseille, CNRS - Luminy, Case 907, 13288 Marseille Cedex 9, France\\ 
Email: qinbo,shanhuany@bao.ac.cn; tilquin@cppm.in2p3.fr}

\begin{abstract}
In a galaxy cluster, galaxies are mostly collisionless particles in recent epoches. 
They resemble collisionless cold dark matter particles in some way. 
Therefore, the spatial distributions of dark matter
and cluster galaxies might be expected to possess similar features in the 
gravitational potential of a cluster. 
Here we use the galaxy distribution in Cluster Cl~0024+17 to probe for the 
ringlike dark matter structure recently discovered by means of strong and weak 
lensing observations. 
The galaxies are taken from the catalog of Czoske et al., 
which contains 650 objects with measured redshifts, of which $\sim$~300 
galaxies have redshifts in the range $0.37<z<0.41$ (and are therefore probable 
cluster members). 
We find that, at about the $3 \sigma$ level, the ringlike structure seen in the 
dark matter measurement is not observed in the projected two-dimensional 
galaxy distribution.
\end{abstract}

\maketitle

\section{Introduction}

The galaxy cluster Cl~0024+17 (at $z\simeq 0.4$) has recently been reported, 
from strong and weak lensing observations, to have a ringlike dark matter 
structure. The radius of the ring was measured to be $\sim 0.4$~Mpc 
(Jee et al. 2007). 
This unique structrue has been suggested as the result of a high-speed 
line-of-sight collision of two massive clusters $\sim 1-2$~Gyr ago. 
Ripples in the mass distribution that were not traced by the distribution 
of baryonic matter (i.e., the intracluster gas) provide strong evidence 
for the existence of dark matter. The same result has also 
been reached for other systems, e.g., the famous bullet cluster 
1E0657-56 (Markevitch et al. 2002; Clowe et al. 2004, 2006), 
clusters Cl~0152-1357 and MS1054-0321 (Jee et al. 2005a, 2005b), 
and some other clusters (Shan et al. 2008). 
Hence, it would be of particular interest to test this unique ringlike 
dark matter structure by other means.

Galaxies in a cluster of galaxies are mostly collisionless particles. Their 
dynamical behavior is quite different from that of the intracluster gas, 
but instead resembles collisionless cold dark matter particles in some way.
Therefore, the galaxy distribution in a cluster could be expected to 
exhibit similar features to the dark matter distribution. 
This has been well demonstrated by the extreme case of the bullet cluster---

While the intracluster gas in Cluster 1E0657-56 has a rather prominent 
offset from the dark matter distribution, which has been used as direct 
evidence against Modified Newtonian Gravity (Milgrom 1983), the cluster 
galaxies, on the other hand, were found to trace the dark matter distribution 
remarkably well (Markevitch et al. 2002; Clowe et al. 2004; 2006). 
Similar results have also been found in other systems such as 
clusters Cl~0152-1357 and MS1054-0321 (Jee et al. 2005a, 2005b).

These findings are the motivations for our attempts to investigate quantitatively 
the galaxy distribution in Cluster Cl~0024+17, in order to probe for its 
reported ringlike structure of dark matter. If the galaxies exhibit similar 
features in their spatial distributions, then they may provide us an 
independent verification of the discovered dark matter ring.

The galaxy cluster Cl~0024+17 has been a target of a number of studies since
its discovery (see Jee et al. 2007 for a historical review, and references therein). Czoske et al. (2001) have made a comprehensive study of the galaxies in 
this cluster, especially their redshift distribution. 
They have compiled a catalog from a wide-field CFHT/WHT survey that contains 
650 objects with measured redshifts, of which 295 galaxies have redshifts in the 
range $0.37<z<0.41$ and are therefore probable cluster members.

Here we use the sample of 295 cluster galaxies from the Czoske et al. (2001) catalog,
and calculate, for each galaxy, the angular separation from the cluster center. 
The radial distribution of the galaxies is then compared with the dark matter distribution in the cluster. 
Throughout this Letter, we adopt the $\Lambda$CDM cosmology with $\Omega_M =0.27$, $\Omega_{\Lambda} =0.73$, and $H_0=71 \; {\rm km} \; {\rm s}^{-1}{\rm Mpc}^{-1}$.

\section{Galaxy Distribution}

The Czoske et al. (2001) catalog provides the redshifts and positions (right 
ascensions and declinations) of 650 objects in the Cl~0024+17 sky area. 
Following Jee et al. (2007), we choose the cluster center to be located at 
the geometric center of the ``dark matter ring'', i.e., 
$\alpha_{2000}\simeq 00^h : 26^m : 35^s.92, \   
\delta_{2000}\simeq 17^{\circ}: 09': 35''.5 $. 
We define the cluster axis to be the line of sight through the cluster center, 
and the radial distance $r$ to be the angular distance of an object from 
the cluster axis. Then we calculate the value of $r$ for each of the 
295 galaxies, from their right ascensions and declinations. For simplicity
$r$ here is in units of arcseconds. 
In the $\Lambda$CDM cosmology with $\Omega_M =0.27$, $\Omega_{\Lambda} =0.73$, 
and $H_0=71 \; {\rm km} \; {\rm s}^{-1}{\rm Mpc}^{-1}$, and at the cluster
distance of $z=0.395$, $1''$ corresponds to an angular diameter 
distance of about $5.28$~kpc.

Numerical simulations have suggested that the collision of the two sub-clusters 
was along the line of sight, and the ``dark matter ring'' was also found to 
be along the line of sight. Thus, the configuration of the cluster is essentially 
two-dimensional and axisymmetric along the line of sight. As a result, 
we are able to treat the galaxy distribution as a two-dimensional problem. 
Assuming isotropy, we calculate the two-dimensional number density of the cluster 
galaxies as a function of $r$ only. 
If the galaxies trace the dark matter distribution, then a concentration
of galaxies near $r=75''$ (corresponding to $\sim 0.4$~Mpc) might be expected.

The results are presented in Figure~1. The two-dimensional number density of the 
295 cluster galaxies is plotted as a function of $r$. 
The histogram shows the {\it observed} galaxy distribution, with
statistical error bars in each bin computed using the Gaussian statistic
approximation as
\begin{equation}
\sigma_i=\frac{\sqrt{n(r_i)}}{\pi(r_i^2-r_{i-1}^2)},
\end{equation}
where $n(r_i)$ is the number of galaxies in each bin, 
and  $r_i-r_{i-1}$ is the width of bin.
We use a data bin of $5''$. 
Following the dark matter profile of Jee et al. (2007), we plot as a 
dotted curve the ``expected'' galaxy distribution, if galaxies follow 
{\it exactly} the dark matter distribution.
The normalization of this curve is chosen such that the curve best-fits the 
galaxy distribution within $r=100''$. The best fit is obtained by minimizing
the total $\chi^2$ for dark mater and galaxy distributions
\begin{equation}
\chi^2=\sum_{i=1}^{n} \left[\frac{\kappa^{DM}(r_i)- \alpha  \kappa^{Gal}(r_i)}
{\sigma}\right]^2,
\end{equation}
where $\alpha$ is the normalization factor, $\kappa^{DM}$ and $\kappa^{Gal}$ 
are the convergence from weak lensing observation and galaxy number
density, respectively.

\begin{figure}[t]
\epsscale{1.1}
\plotone{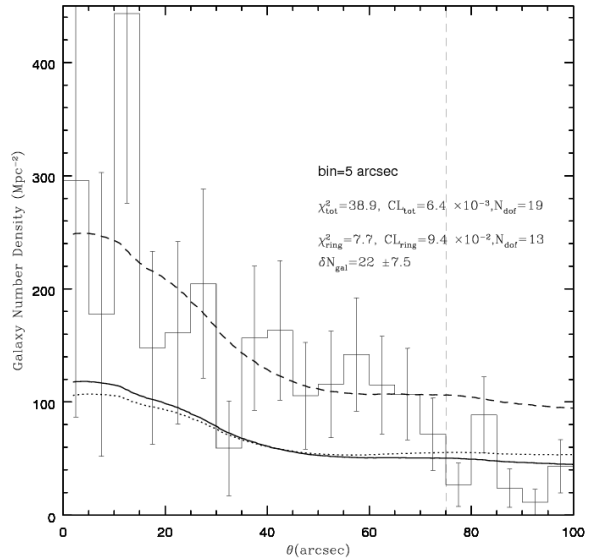}
\caption{2-dimensional galaxy number density in the cluster Cl~0024+17, as a 
function of the angular distance from the cluster center. 
The histogram shows the observed galaxy distribution, while the 3 curves 
are the theoretically ``predicted'' galaxy distributions if they follow 
the dark matter distribution reported by Jee et al. (2007).
The dotted curve shows the result before the completeness correction, and
the solid curve is the galaxy distribution after the completeness correction.
The normalizations of the two curves are chosen such that both curves 
best-fit the observed galaxy distribution. 
The dashed curve is the best fit galaxy distribution for regions $r \le 75''$.
The vertical dashed line marks the position of the ``dark matter ring'' 
at $r=75''$.
We choose the cluster center to be the geometric center of the ring.}
\end{figure}

The position of the ring is marked by the vertical dashed line,
i.e., at a radial distance of $r=75''$. Figure~1 displays only the galaxy 
distribution within $r=100''$ (or $\sim 1$~Mpc), in accordance with the 
dark matter profile range of Jee et al. (2007). 

Indeed, the dotted curve in Figure~1 cannot be used directly for comparison  
with the observed galaxy distribution, due to the selection biases caused 
by the completeness of the Czoske et al (2001) catalog.  
The completeness of the sample varies with radius, 
from $>80\%$ at the cluster center to $<50\%$ in the outer regions. 
The contour lines in Figure~2 show the completeness variation (see also 
Fig.~7 in Czoske et al. 2001). 
Therefore, a completeness correction is needed to mimic a more 
``realistic'' galaxy distribution, in order to compare with the 
actually observed galaxy distribution.

To do this, we obtain from the two dimensional completeness contours an 
averaged completeness that varies with the cluster radius only (one dimensional). 
To compute the average completeness, we simulate 100,000 galaxies for 
$r \le 100''$, assuming galaxies follow dark matter. 
Each galaxy is then selected according to the completeness value at its position. 
The average radial completeness is computed as the observed number of galaxies 
in an $r$ bin divided by the number of simulated galaxies in the corresponding bin.
Our expected galaxy distribution is then corrected 
from the dotted curve to the thick solid curve in Figure~1. 
The result of the completeness correction is that the bump in 
the dotted curve near $r=75''$ has been slightly suppressed.

The new normalization factor $\alpha$ has been fitted by minimizing the
$\chi^2$ for the whole region of $r\le 100''$. 
The minimum is found to be $\chi^2=38.9$, 
with a number of degrees of freedom $N_{\rm dof}=19$. The confidence level of the fit
is CL$=0.64\%$. Such a low confidence level suggests an 
inconsistency between the dark matter and galaxy distributions.
The dashed curve is obtained by fitting the normalization factor 
for $r\le r_{\rm ring}=75''$. The $\chi^2$ is found to be equal to
$7.7$, with a number of degree of freedom $N_{\rm dof}=13$ and a 
confidence level CL$=9.4\%$, indicating a much better agreement
between both distributions.
We then estimate, for the dashed curve, the number of galaxies 
$\delta N_{\rm gal}$ in the $r>r_{\rm ring}$ regions in Figure~1, 
and compare it to observations.   
We find that $\delta N_{\rm gal}= 22.0 \pm 7.5$,
which suggests that, at a level of about $3\sigma$, the newly discovered 
ringlike structure of dark matter in Cl~0024+17 is not mimiced by the 
distribution of the galaxies in the cluster. Such a result is quite different 
from the case of the bullet cluster (Markevitch et al. 2002;
Clowe et al. 2004, 2006), or clusters 
Cl~0152-1357 and MS1054-0321 (Jee et al. 2005a, 2005b), where the galaxy 
distributions were found to trace the dark matter distributions well.

To estimate more precisely the confidence level of our probe of the ringlike
dark matter structure using the observed galaxies, we have used a simulation of 
the cluster survey. For this purpose we extract the completeness curves from 
the Czoske et al. (2001) catalog. Galaxies have then been isotropically 
simulated according to the radial dark matter distribution and 
passed through completeness using a simple Metropolis method. 
Figure~2 shows a given Monte Carlo realization of a simulated survey. 

Each simulated experiment is normalized with the observed 72 galaxies
within the dark matter ring randomized using the Poisson statistics. 
An experiment is accepted if the number
of simulated galaxies inside the ring is smaller than or equal to the
number of actually observed galaxies (17 galaxies).
The confidence level exclusion is computed as one minus the number of accepted 
experiments divided by the total number of simulated experiments.  
Using 10,000 simulated experiments we find a
confidence level exclusion of $99.6\%$, which confirms our previous analysis.

\begin{figure}
\epsscale{1.}
\includegraphics[angle=0,width=8.9cm]{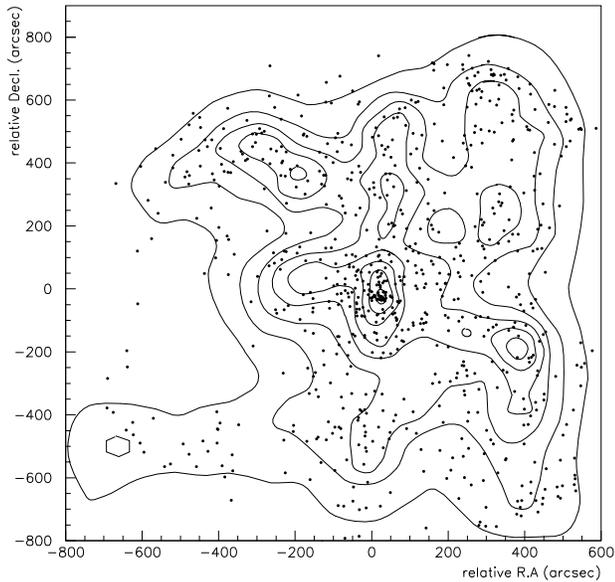}
\caption{Simulated galaxies from a given Monte Carlo realization according to
the dark matter distribution. Contour lines (from Czoske et al. 2001) are 
the completeness of the sample which are spaced in $10\%$ step, 
with the innermost contour being $80\%$.}
\end{figure}

\section{Conclusion and Discussion}

Unlike the intracluster gas, galaxies in a cluster of galaxies are  
mostly collisionless particles. Their dynamical properties resemble the
properties of dark matter to some extent. Therefore, under the same gravitational 
potential of a cluster, one may expect to find similar features 
between dark matter and galaxy distributions. In this respect, galaxy 
distribution might be used to probe dark matter distribution, 
especially for rich clusters.

Indeed, the bump at $r \sim 75''$ in the mass density profile of Jee et al. (2007), 
which gives rise to the reported dark matter ring, is a quite remarkable 
feature. An increase in galaxy number density at $r \sim 75''$ should be 
expected in Figure~1, if galaxies trace dark matter. 
Unfortunately, at the $3\sigma$ level, cluster Cl~0024+17 does not appear 
to exhibit the level of consistency between dark matter and galaxy 
distributions that have been found in other clusters such as  
the Bullet Cluster, Cl~0152-1357, and MS1054-0321.

The reason for this inconsistency could be that the sample of 295 galaxies is 
not large enough for a stringent constraint, and the galaxies are too sparsely 
distributed within the radius of $\sim 1$~Mpc. We have realized this disadvantage. 
Fortunately, the axisymmetrical configuration of the system along the line of 
sight allows us to plot the surface density of the galaxies as a function of 
radius $r$ only. This has greatly increased the number of galaxies in each bin 
of our Figure~1, and hence improved the quality of our data.  
We have also tried different bin sizes, but reached similar results. 
Our simulation has been found to be insensitive to the above effects   
and confirms our $3\sigma$ confidence level. 


It should be noted that the completeness given by Czoske et al (2001)
has presumably been smoothed over an area comparable to or larger than 
the width of the dark matter ring. This could reduce the contrast in 
galaxy number density that we derived from the completeness correction. 
However, if the completeness does not vary drastically around the 
position of the ring, this may not lead to a significant change 
in our result.

The inconsistency could also be because, as argued by Jee et al. (2007), 
the density contrast in the ringlike structure is low. 
But however, compared to a pure NFW distribution (Navarro, Frenk, \& White 1997) 
where the dark matter density decreases rapidly and monotonically with radius, 
the bump at $r \sim 75''$ as shown in the dark matter density profile of
Jee et al. (2007) is still quite prominent. 
This prominent feature, if it really exists, might be ``seen'' 
if we had enough luminous ``tracers'' of dark matter (e.g., 
the cluster galaxies). 
Interestingly, a recent study has shown that ringlike artefacts 
could be produced in lens reconstructions (Liesenborgs et al 2008).

Nevertheless, as has been demonstrated in this Letter, the galaxy spatial 
distribution in a cluster of galaxies could provide a useful and 
complementary tool for our study of the dark matter distribution, 
due to the similarities in the collisionless nature of cluster galaxies and 
cold dark matter particles. This would be of particular interest
for rich clusters of galaxies.

\acknowledgements 
We thank Prasenjit Saha, Charling Tao, Xiang-Ping Wu, Christopher Ke-Shih Young, 
Pengjie Zhang, and Xinmin Zhang for discussions 
and an anonymous referee for helpful comments.
This work was supported by a CAS grant KJCX3-SYW-N2.

\end{document}